\documentclass[conference,a4paper]{IEEEtran}
\usepackage[left=1.62cm,right=1.62cm,top=1.9cm]{geometry}
\usepackage{balance}
\usepackage{float}

\usepackage{graphicx}
\usepackage{subfig}

\usepackage{standard}
\usepackage{etex}

\usepackage{rotate}
\usepackage{algorithmic}
\usepackage{algorithm}
\usepackage{float}
\usepackage{tikz}
\usepackage{refcount}

\usepackage{pgf,tikz}
\usepackage{pgfplots}
\usepackage{pgfplotstable}
\usepackage{bbm}
\usepackage{adjustbox}
\usetikzlibrary{calc}
\usepackage{relsize}
\usepackage{ifthen}

\usetikzlibrary{calc,fit,arrows,plotmarks,intersections,patterns,shapes,decorations,positioning, backgrounds, matrix}

\usepackage[numbers,sort&compress]{natbib}
\usepackage[ngerman, english]{babel}

\usetikzlibrary{arrows.meta}
\edef\wdArrowLength{2}
\tikzset{>={Latex[width=1.5mm,length=\wdArrowLength mm]}}
\usepackage{graphicx}

\usepackage{cancel}
\usepackage{mathtools}
\usepackage{shadethm}
\usepackage{empheq}
\usepackage{skull}
\usepgfplotslibrary{units}
\usepackage{calc}
\usepackage{nicefrac}
\usepackage{fancybox}
\usepackage{psfrag}
\usepackage{dsfont}

\title{Show Me the Way: Real-Time Tracking of Wireless Mobile Users with UWB-Enabled RIS
\thanks{This work was supported in part by the German Federal Ministry of Education and Research (BMBF) in the course of the 6GEM Research Hub under grant 16KISK03 and in part by 6G-ANNA under grant
16KISK095. K. Weinberger, S. Tewes and A. Sezgin are with the Ruhr-Universit\"at Bochum, Germany Email:\{kevin.weinberger, simon.tewes, aydin.sezgin\}@rub.de.}} 
\author{Kevin Weinberger, Simon Tewes, Aydin Sezgin}
\date{\today}

\usepackage{graphicx}
\usepackage{placeins}

\usepackage{booktabs}
\usepackage{marvosym}
\usepackage{multirow}

\usepackage{latexsym, amsmath, amssymb, amsfonts, upgreek}

\usepackage[nolist]{acronym}

\usepackage{tikz}
\usetikzlibrary{calc,arrows,positioning,decorations,shadows,shapes,fadings,matrix}
\tikzset{>=latex'}
\tikzset{semithick}

\makeatletter
\providecommand{\IfElsePackageLoaded}[3]{\@ifpackageloaded{#1}{#2}{#3}}
\makeatother

\makeatletter

\def\tikz@delimiter#1#2#3#4#5#6#7#8{%
	\bgroup
		\pgfextra{\let\tikz@save@last@fig@name=\tikz@last@fig@name}%
		node[outer sep=0pt,inner sep=0pt,draw=none,fill=none,anchor=#1,at=(\tikz@last@fig@name.#2),#3]
		{%
			{\nullfont\pgf@process{\pgfpointdiff{\pgfpointanchor{\tikz@last@fig@name}{#4}}{\pgfpointanchor{\tikz@last@fig@name}{#5}}}}%
			\delimitershortfall\z@
			\resizebox*{!}{#8}{$\left#6\vcenter{\hrule height .5#8 depth .5#8 width0pt}\right#7$}%
		}
		\pgfextra{\global\let\tikz@last@fig@name=\tikz@save@last@fig@name}%
	\egroup%
}

\tikzset{hexagon/.code={
	\draw (0,2) -- (-4,0) -- (0,-2) -- (4,0) -- (0,2);
}}

\tikzset{phone/.code={
   \node [rectangle,rounded corners=1.5pt,draw,minimum height=0.6cm, minimum width=0.35cm] at (0,0){};
   \node [rectangle,rounded corners=1.5pt,draw,minimum height=0.5cm, minimum width=0.3cm] at (0,0){};
}}

\makeatletter
\def\cantox@vector#1#2#3#4#5#6#7#8{%
  \dimen@.5\p@
  \setbox\z@\vbox{\boxmaxdepth.5\p@
   \hbox{\kern-1.2\p@\kern#1\dimen@$#7{#8}\m@th$}}%
  \ifx\canto@fil\hidewidth  \wd\z@\z@ \else \kern-#6\unitlength \fi
  \ooalign{%
    \canto@fil$\m@th \CancelColor
    \vcenter{\hbox{\dimen@#6\unitlength \kern\dimen@
      \multiply\dimen@#4\divide\dimen@#3 \vrule\@depth\dimen@\@width\z@
      \vector(#3,-#4){#5}%
    }}_{\raise-#2\dimen@\copy\z@\kern-\scriptspace}$%
    \canto@fil \cr
    \hfil \box\@tempboxa \kern\wd\z@ \hfil \cr}}
\def\bcancelto#1#2{\let\canto@vector\cantox@vector\cancelto{#1}{#2}}
\makeatother

\newcommand{\ifthen}[2]{\ifthenelse{#1}{#2}{}}

\definecolor{myblue1}{rgb}{0,0,255}
\definecolor{myblue2}{rgb}{65,105,225}
\definecolor{myblue3}{rgb}{70,130,180}
\definecolor{myblue4}{rgb}{176,196,222}

\newcommand{\mytilde}{{\raise.17ex\hbox{$\scriptstyle\mathtt{\sim}$}}}
\newcommand{\naive}{}
\def\naive/{na\"{\i}ve}

\newcommand{\executeiffilenewer}[3]{%
\ifnum\pdfstrcmp{\pdffilemoddate{#1}}%
{\pdffilemoddate{#2}}>0%
{\immediate\write18{#3}}\fi%
}

\pgfkeys{/pgf/fpu}
\pgfmathparse{16383+1}

\pgfkeys{/pgf/fpu=false}

\IEEEoverridecommandlockouts

\begin{document}
\bstctlcite{IEEEexample:BSTcontrol}

\maketitle

\begin{abstract}

  The integration of Reconfigurable Intelligent Surfaces (RIS) in 6G wireless networks offers unprecedented control over communication environments. However, identifying optimal configurations within practical constraints remains a significant challenge. This becomes especially pronounced, when the user is mobile and the configurations need to be deployed in real time. Leveraging Ultra-Wideband (UWB) as localization technique, we capture and analyze real-time movements of a user within the RIS-enabled indoor environment. Given this information about the system's geometry, a model-based optimization is utilized, which enables real-time beam steering of the RIS towards the user. However, practical limitations of UWB modules lead to fluctuating UWB estimates, causing the RIS beam to occasionally miss the tracked user. The methodologies proposed in this work aim to increase the compatibility between these two systems. To this end, we provide two key solutions: beam splitting for obtaining more robust RIS configurations and UWB estimation correction for reducing the variations in the UWB data. Through comprehensive theoretical and experimental evaluations in both stationary and mobile scenarios, the effectiveness of the proposed techniques is demonstrated. When combined, the proposed methods improve worst-case tracking performance by a significant 17.5dB compared to the conventional approach.
\end{abstract}
	\vspace{-0.1cm}
\section{Introduction}
Reconfigurable Intelligent Surfaces (RISs) stand as a transformative force in shaping the landscape of future sixth-generation (6G) wireless communication networks \cite{Renzo2}. Comprising multiple individually controllable reflecting elements, these surfaces offer unparalleled control over communication environments, promising advancements in capacity, resilience \cite{outageResilience}, and energy efficiency \cite{EEJourn}. However, determining optimal configurations in real-time and under practical constraints poses a significant challenge. Utilizing state-of-the-art optimization methods for these scenarios becomes infeasible due to their complexity and the time required to solve them. Moreover, determining the individual channel estimates for each reflect element, which are required to obtain optimal configurations, also becomes a challenge, particularly in time-sensitive applications \cite{RIS_time}.

This study aims to address these problems by deploying a  Ultra-Wideband (UWB)-equipped RIS. Leveraging UWB as a localization technique, the RIS is able to capture and analyze real-time movements of the user within the RIS-enabled environment. This is different from the current proposals in the literature, where UWB-based RIS channel estimations are based on UWB transmissions within the same frequency band over the RIS link itself \cite{RIS_time,RIS_UWB_chanEst,RIS_UWB_est}. Our proposed approach, however, leverages UWB in a disjoint frequency band to estimate the distance and angle between the RIS and the user. Given this data,  we are able to determine the current geometry of the system, under the assumption that the location of the transmitter is fixed and known. By utilizing a model-based optimization approach that also considers practical limitations, we are able to calculate and deploy the optimal RIS configurations in real time \cite{KevChanMod}. Nevertheless, this also makes the RIS's tracking performance highly reliant on the quality of the UWB.

For this reason this work proposes two techniques, which augment the interplay between the RIS and UWB-system. The first approach splits the RIS in multiple sub-surfaces, with each surface adjusted to direct a different beam around the current UWB estimate. This effectively widens the overall RIS reflected beam to compensate for UWB estimation errors at the cost of beam intensity. The second technique aims to correct the UWB estimations directly. By proposing a correction procedure, inspired by the momentum-based gradient descent method \cite{grad_desc_mom}, the fluctuations of UWB estimates can be drastically reduced.
The effectiveness of both methods is validated using numerical simulations and experimental measurements.

\section{System and Channel Model}\label{sec:sysmod}
The system model described in this paper assumes a single-antenna transmitter (Tx) directing signals toward a reconfigurable intelligent surface (RIS), comprised of 4x3 modules with \( M=256 \) reflective elements each, mirroring the prototype used in our experiments as shown in Fig. \ref{fig:Setup}. The reflected signals are then received by a single-antenna receiver (Rx). With the RIS's capability to manipulate phase shifts in the reflections, the effective channel between Tx-RIS-Rx can be expressed as
\vspace{-0.15cm}
\begin{align}\label{eq:heff}
    h^\mathsf{eff} = \sqrt{G_{T}}\sqrt{G_{R}} \sum_{m=1}^{12M} h_m \theta_m g_m,\\[-18pt]\nonumber
\end{align}
where \( h_m \) (\( g_m \)) represents the channel coefficient between the transmitter (receiver) and the \( m \)-th RIS element. $G_{T}$ and $G_{R}$ are the antenna gains at the transmitter and receiver, respectively. The phase shift induced on the reflection by the \( m \)-th RIS element is denoted as \( \theta_m = A_m(\varphi_m)e^{j\varphi_m} \), where \( \varphi_m \) signifies the adjustable phase and \( A_m \in [0,1] \) denotes the phase-dependent reflect amplitude.

Considering the above, the cascaded channel between Tx-RIS-Rx over the \( m \)-th RIS element is formulated as\vspace{-0.1cm}
\begin{align}
     h^\mathsf{casc}_m = h_m g_m,\\[-18pt]\nonumber
\end{align}
where \( h_m g_m \) represents the resulting cascaded channel component. To analytically determine these components for each reflect element \( m \), we utilize the setup's geometry. We define \( d^h_m \) (\( d^g_m \)) as the distance between the transmitter (receiver) and the \( m \)-th RIS element. As per \cite{goldsmith2005wireless}, the cascaded channel between Tx-RIS-Rx over the \( m \)-th RIS element is given by:
\begin{align}\label{eq:chanModel}
     h^\mathsf{casc}_m = h_m g_m = \left[ \frac{c}{4\pi f d^h_m} e^{j\frac{2\pi}{\lambda}d^h_m} \right] \left[ \frac{c}{4\pi f d^g_m} e^{j\frac{2\pi}{\lambda}d^g_m} \right],
\end{align}
where \( c \) represents the speed of light, \( f \) is the frequency and \( \lambda \) signifies the signal wavelength.

\section{Experimental Validation}\label{sec:exp}
\subsection{Experimental Setup}\label{subsec:expSetup}

\begin{figure}[ht]
	\centering
	\includegraphics[width=0.74\linewidth]{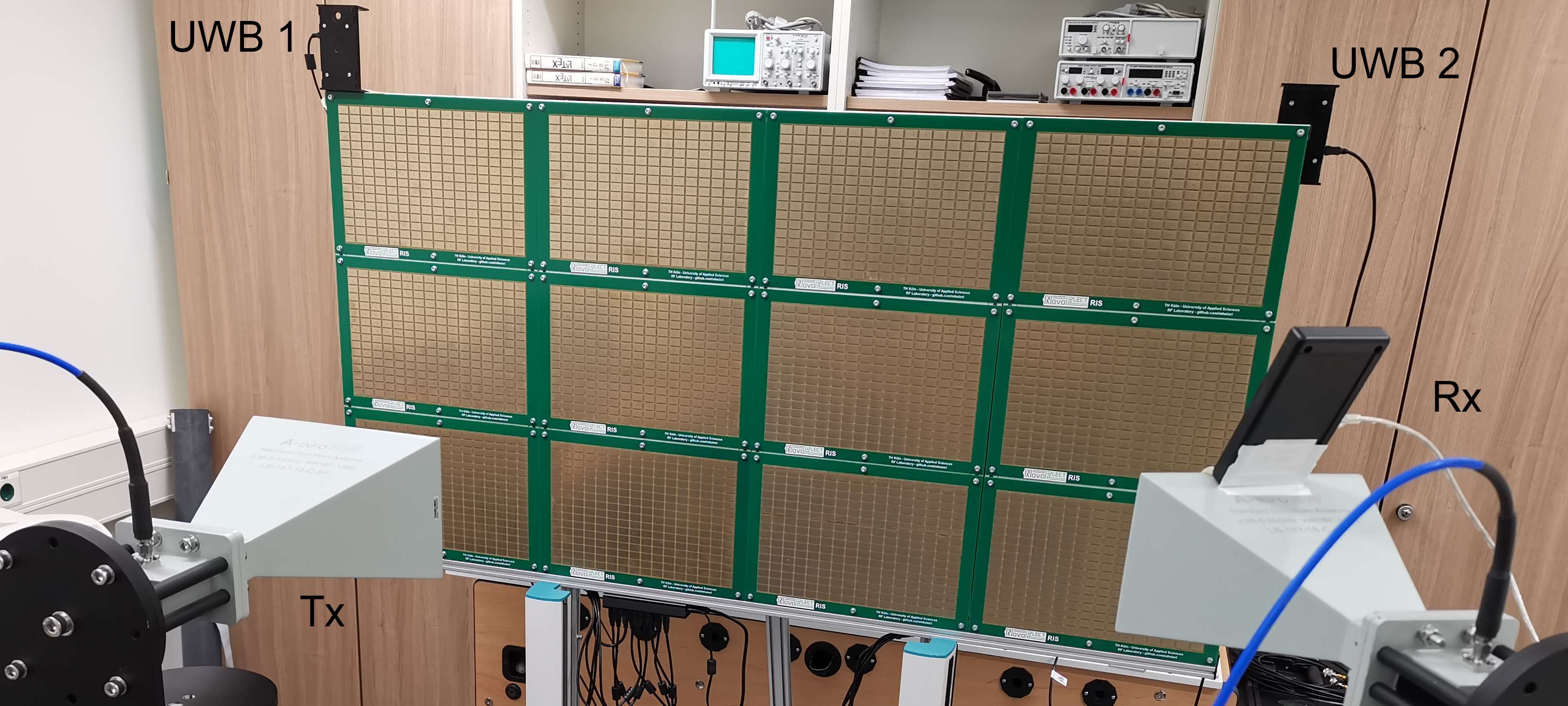}
	\caption{RIS and antennas in an indoor office environment.}
	\label{fig:Setup}
\end{figure}

To validate the precision of the tracking performance, we implemented a scenario using an RIS prototype operating in the 5GHz band. A visual representation of the setup is depicted in Fig.~\ref{fig:Setup}. Each RIS prototype module measures 360 by 247 mm and is detailed further in \cite{ISAPpaperMH}. In our setup, we arranged 12 RIS modules in a 4x3 configuration within an office environment to expand the near-field of the surface. Each RIS module consists of 256 elements arranged in a 16 by 16 grid, with each element featuring an RF switch for phase shifting. The surface functions as a binary-switching surface with a $180^\circ$ phase shift at the designed carrier frequency of $f = 5.53$GHz, corresponding to a wavelength of $\lambda=5.42$cm.

To measure the path loss over the RIS, we utilized a vector network analyzer (VNA) of type PicoVNA 106 from PicoTech. The S21 parameter was evaluated over a frequency span of 200 MHz, ranging from 5.4 GHz to 5.6 GHz, with a resolution bandwidth of 10 kHz. Each port of the VNA was connected to a directional horn antenna of type LB-187-15-C-SF from A-Info, with a minimum gain of 16.89 dBi within the chosen frequency span. The transmit power was set to 0 dBm.

To capture the geometry of transmit and receive antennas relative to the RIS, an UWB localization system based on Decawave DWM1000 modules was employed. The modules operate at a carrier frequency of 6.5 GHz with a bandwidth of 500MHz. Two modules (anchors) were mounted on the RIS, one at the upper right and the other at the upper left corner of the RIS (Fig. \ref{fig:Setup}). The distance between these two modules was 1.41m. Another module was attached to the mobile receiving antenna. Using Two-Way-Ranging, the distance between the receiving antenna and the two modules on the RIS was continuously measured. By utilizing the two distance measurements and the known distance between the anchor modules, the Angle of Arrival (AoA) and the distance to the center of the RIS were determined. These AoA and distance measurements served as input for the RIS optimization presented in this paper.

To ensure synchronization among the RIS, VNA, and UWB modules, all systems were connected to a host computer and controlled via MATLAB. This ensured that RIS switching, VNA measurements, and UWB measurements were consistently performed in the correct sequence.



\subsection{Model-based Optimization}
To enable real-time optimization of the RIS configurations for maximizing the received signal strength, we utilize the channel model specified in section \ref{sec:sysmod}. Based on the estimated position data obtained by the UWB-System, we can generate the current geometry of the system to determine the channels between the moving receive antenna and the reflecting elements of the RIS. Thus, given the channel values, we are able to compute the optimal RIS configuration and round the values to the closest possible switch state. Given the absence of a direct link between Tx and Rx, we account for various phase values in the effective channel to optimize the RIS state for achieving the most efficient configuration.

In order to guarantee real-time capability, we compute the optimal RIS configuration analytically, expressed as $\varphi_m^*(C_t) = C_t - \varphi'_m$. Here, $C_t$ denotes an arbitrary value for the desired phase at the receiver, due to the absence of a direct Tx-Rx link, and $\varphi'_m = \frac{2\pi}{\lambda^*}(d_m^h + d_m^g)$. Although the flexibility in selecting $C_t$ allows us to iterate through any number of evenly-distributed phase values $C_t$, we reduce the maximum number of allowed values to $T = 8$. This achieves a balanced tradeoff between the quality of the RIS-enabled link and the time needed in order to determine the optimal configuration. After determining the analytical solution for the $T$ phase values, we round the continuous phase shifts at each reflecting element to the nearest possible binary switching state of the RIS prototype. For these prototype-deployable RIS configurations, we choose the best performing one by assessing and comparing the simulated performances of the resulting RIS-facilitated links. This process can be represented mathematically as \vspace{-0.4cm}

\begin{align}
h_t^\mathsf{eff} = &\hspace{-0.1cm} \sum_{m=1}^{12M} {h}^\mathsf{casc}_m A_m(\text{rd}(\varphi_m^*(C_t))) \e^{j \text{rd}(\varphi_m^*(C_t))},\,  \forall C_t \in \mathsf{\Theta},\label{optEq}\\[-8pt]
 \text{with}&  \quad A_m(\tau) = \begin{cases} $0.5012 (-3\text{dB})$ , \quad\text{if } \tau = \pi \\ 1 , \quad\quad\quad\quad\quad\,\, \,\,\,\,\:\text{otherwise}\end{cases}\ \,\,\,\, ,\\
 &\quad\,\,\,\, \text{rd}(\tau) = \begin{cases} \pi , \quad\text{if } \frac{\pi}{2} \leq\tau < \frac{3\pi}{2} \\ 0, \quad\text{otherwise}\end{cases}\quad\quad\quad\,\,\,\,\,\,  , \label{rdFun}\\[-1pt]
 &\quad\quad\quad\mathsf{\Theta} = \{0, \frac{\pi}{180}, \frac{2\pi}{180}, \frac{3\pi}{180}, \ldots, 2\pi - \frac{\pi}{180}\} ,\label{Theta} \\[-13pt] \nonumber
\end{align}	
where an additional attenuation of 3dB to the reflected path of reflecting element $m$  is assumed, if it is turned on. This attenuation is considered in the simulations in order to capture the hardware limitations of the utilized RIS prototype~\cite{ISAPpaperMH}.

\subsection{Beam Splitting}

We intend to capture uncertainties in the channel model and UWB data in order to increase the robustness of the tracking performance by splitting the RIS-reflected beam. More precisely, we divide the RIS into sub-surfaces and generate multiple individual beams towards the receive antenna. Due to the size of the surface, we assume the transmit and receive antenna are both in the near-field of the surface. Consequently, the optimization process results in the surface acting as a lens, which generates a focal point in space. This means, that errors, e.g., during distance and angle estimation by the UWB-System, can result in a vastly reduced channel quality, especially when applying the optimized RIS configurations in practice. To counteract this problem, we introduce two beamsplitting methods, which are also analyzed simulatively. The first method is the angle-splitting method (ASM), for which we define a splitting factor $\tilde{\nu}^\mathsf{UWB}$ for the UWB-estimated angle $\nu^\mathsf{UWB}$. Instead of optimizing the whole RIS for this point, we generate multiple points adjacent to the estimated position, for example $\nu_{\{+,-\}}^\mathsf{split} = \nu^\mathsf{UWB} \pm \frac{\tilde{\nu}^\mathsf{UWB}}{2}$ for the two beam case. Further, we split the RIS horizontally into multiple sections and optimize each section individually for one of the points, i.e., $\nu_{\{+,-\}}^\mathsf{split}$ for the two beam case, using (\ref{optEq})-(\ref{Theta}). The second method is the distance-splitting method (DSM), in which the principle of the ASM is applied to the estimated distance of the UWB-system, i.e., $d_{\{+,-\}}^\mathsf{split} = d^\mathsf{UWB} \pm \frac{{\tilde{d}^\mathsf{UWB}}}{2}$ for the two beam case, where ${\tilde{d}^\mathsf{UWB}}$ is the beam splitting factor.

\subsection{Phase Matching}
When utilizing the beam-splitting methods, the goal is to generate new beams at the RIS that are optimized for points close to the respective UWB estimations. As a result the new beams might result in a deterioration with regards to the channel quality at the estimated point $\{d^{\mathsf{UWB}},\nu^\mathsf{UWB}\}$, but will offer a better area of coverage around it. In order to reduce the deterioration of the channel quality at $\{d^{\mathsf{UWB}},\nu^\mathsf{UWB}\}$ during the beam-splitting methods, we select the combination of phase values $C_t$ for all individual beams such that the channel quality at the estimated point $\{d^{\mathsf{UWB}},\nu^\mathsf{UWB}\}$ is maximized, i.e., the beams interfere with each other constructively at $\{d^{\mathsf{UWB}},\nu^\mathsf{UWB}\}$.

\begin{figure}
\centering
    \includegraphics[width=.75\linewidth]{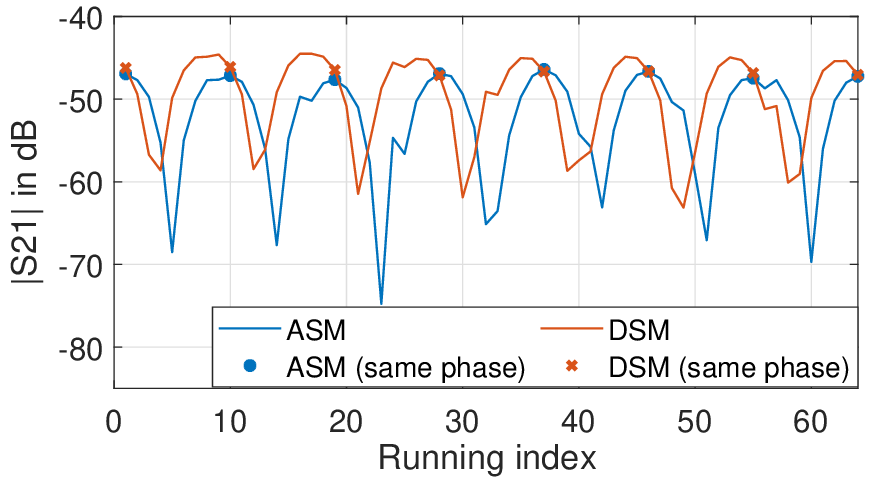}
    \caption{The logarithmic magnitude of the channel values at $\{d^{\mathsf{UWB}},\nu^\mathsf{UWB}\}$ for each possible combination during phase matching procedure using the DSM with $\tilde{d}^\mathsf{UWB}=0.1$ m and ASM with $\tilde{\nu}^\mathsf{UWB}=2.5^\circ$ and $T=8$.}
    \label{fig:phasematch}
\end{figure}

To simplify the analysis, we assume that both the ASM and DSM utilize the two beam split. Fig. \ref{fig:phasematch} depicts the channel quality at $\{d^{\mathsf{UWB}},\nu^\mathsf{UWB}\}$ for each of the $T^2$ phase value combinations of the newly optimized beams for the ASM and DSM, respectively. The figure shows that for the ASM the optimal phase combinations are mostly the ones that have a similar phase, i.e., $C_{t,+} \approx C_{t,-}$. The figure also shows, that a bad selection of phase values for the individual beams can result in a significant reduction of the channel quality. This becomes especially pronounced when beams with opposing phase values are chosen, which results in destructive interference at $\{d^{\mathsf{UWB}},\nu^\mathsf{UWB}\}$. For the case of the DSM, the figure depicts a different behavior. Here, non-similar phases for the new beams become more effective. In addition, a bad selection of phase values during DSM does not reduce the channel quality as much as in ASM.

\subsection{UWB Estimation Correction}
Due to the imprecise UWB estimation, the measured values for distance and angle to the Rx antenna are fluctuating, even for a static scenario. In order to reduce these variations, we propose an approach that mimics the functionality of the momentum-based gradient descent method \cite{grad_desc_mom}. To this end, we define the following equation for the gradient-based values
\begin{align}
  &\hat{q}_{k+1} = \begin{cases}{q}_{1}, \hspace{4.7cm}\text{if } k=\{0,1\}, \label{x}\\ \hat{q}_{k} + \beta(q_k\hspace{-0.05cm}-q_{k-1}) + (1-\beta)(\hat{q}_k-\hat{q}_{k-1}), \text{otherw.}\end{cases}
\end{align}
where $q\in\{\nu^\mathsf{UWB},d^\mathsf{UWB}\}$ are the UWB-estimated distances and angles, $\hat{q}$ are the respective gradient-based values, $\beta\in [0,1]$ is the momentum parameter and $k$ the current sample number. By using this equation, the subsequent gradient-based value $\hat{q}_{k+1}$ is calculated from the current gradient-based value $\hat{q}_k$ and a weighted sum of the gradients from the current and last UWB as well as gradient-based points. Here, the weighting of $\beta$ determines, how much the individual gradients impact the calculation of the next corrected point, i.e., $\beta = 1$ implies that only the gradient of the UWB values is considered during calculation. Due to the fact, that the gradients can sometimes result in overestimations during the movement of the Rx antenna, we define the corrected values $\overline{q}_k$ as the mean of the current gradient-based and UWB value $\overline{q}_k = \frac{q_k + \hat{q}_k}{2}$.  For the remainder of this work we set $\beta_\nu=0.55$ and $\beta_d=0.3$, which were determined based on observations during the measurement campaign.

\section{Simulation Results}
\subsection{Analysis of Beamsplitting and Phase Matching}
\begin{figure}
\centering
    \includegraphics[width=.65\linewidth]{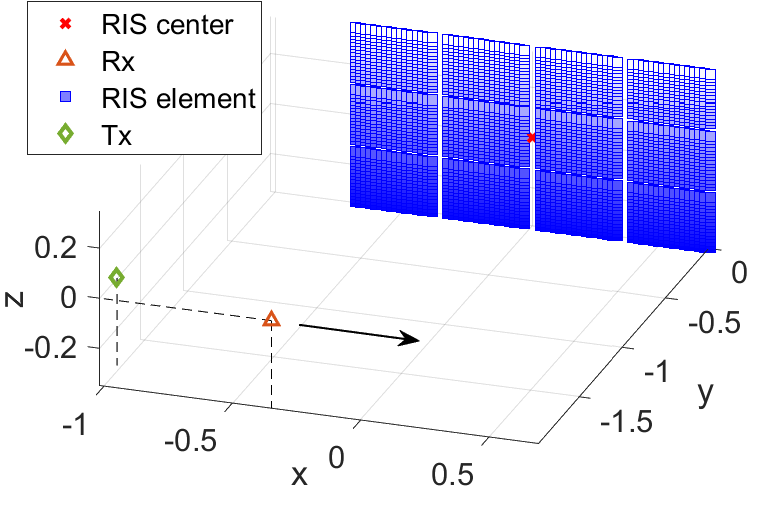}
    \caption{Virtual representation of the experimental setup. The arrow indicates the direction of the trajectory during the simulations.}
    \label{sysmod1}
\end{figure}

In this section, the effects and differences of beamsplitting and phase matching are studied simulatively. To this end, we assume a setup, which coincides with the one described in section \ref{subsec:expSetup} and is visualized in Fig. \ref{sysmod1}. The Tx antenna is located at a radial distance of 2m from the center of the RIS and is positioned so that the angle of incidence at the RIS center is $-30^\circ$. The Rx antenna is initially placed at $d^\mathsf{Rx} = 2$m and $\nu^\mathsf{Rx} = -10^\circ$, with both antennas situated at the same height as the RIS center. The Rx antenna follows a trajectory of $0.05$m/s radially and $5^\circ$/s angularly. Throughout this movement, we configure the RIS using the ASM and DSM, as well as the conventional method, assuming perfect estimations from the UWB-system. We set $\tilde{d}^\mathsf{UWB}=0.1$ m for DSM and $\tilde{\nu}^\mathsf{UWB}=2.5^\circ$ for ASM, unless otherwise specified. Notably, these values result in new beams generated by the ASM and DSM, which deviate from the UWB-estimated point by approximately 5cm.

\subsubsection{Beamsplitting}
\begin{figure}
\centering
    \includegraphics[width=.95\linewidth]{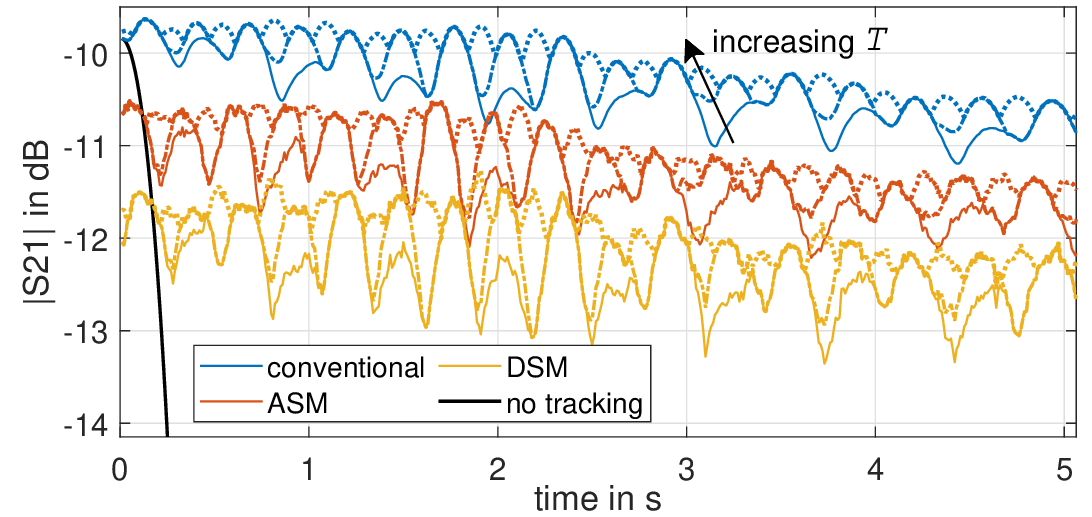}
    \caption{The logarithmic magnitude of the channel values at $\{d^{\mathsf{UWB}},\nu^\mathsf{UWB}\}$ simulated for the conventional optimization as well as using the ASM with $\tilde{d}^\mathsf{UWB}=0.1$ m and DSM with $\tilde{\nu}^\mathsf{UWB}=2.5^\circ$ for $T=\{1,2,4\}$.}
    \label{fig:split1}
\end{figure}

Fig. \ref{fig:split1} shows the logarithmic magnitude of the effective channel values for the conventional and the beamsplitting approaches. First, we focus on the case $T=1$, in which the RIS is optimized towards one channel phase value only.  Naturally, the conventional method outperforms the beamsplitting methods due to the assumption that the UWB-estimates are perfect. Compared to the conventional method,  ASM results in loss of about 1dB and DSM in a loss of about 2dB. The increased loss observed for the DSM is caused by destructive interference of the reflected signal at the UWB-estimated point.

When increasing the number of phase values to $T=2$, the figure illustrates that the performance for all methods can improve by approximately 1dB at specific points. At all other points, it is evident that the performance remains unchanged with the addition of more channel phases, rather than declining. The benefit of adding more channel phase values comes from the quantization of the phase shifts. Specifically, due to the quantization, some channel phases can become unfavorable at certain geometries. However, these same geometries may also align favorably with other phase values. This becomes apparent when comparing the curves of the same method for different values of $T$. By increasing the number of phase values, unfavorable alignments are compensated for, as there are more options to choose from.

\subsubsection{Phase Matching}
\begin{figure}
\centering
    \includegraphics[width=.95\linewidth]{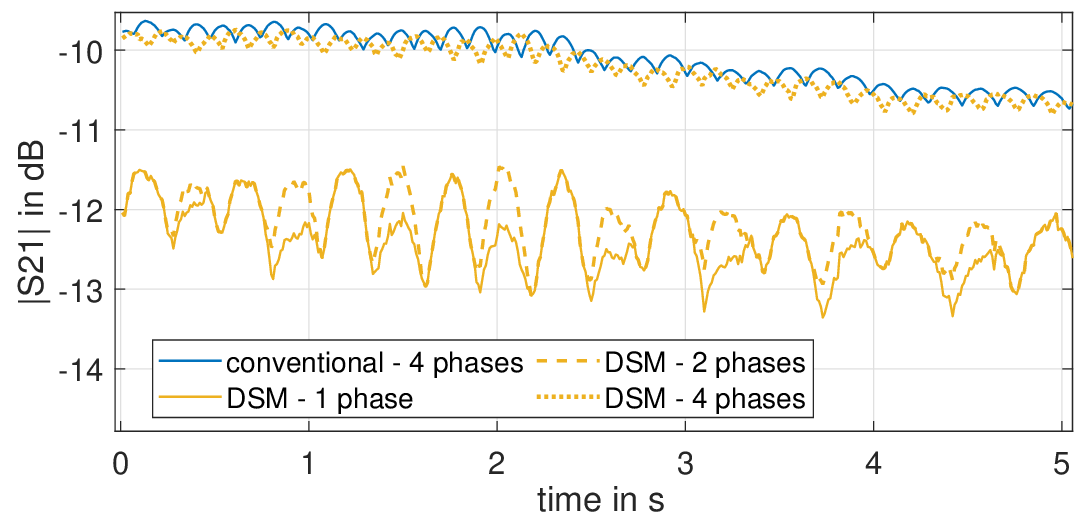}
    \caption{The logarithmic magnitude of the channel values at $\{d^{\mathsf{UWB}},\nu^\mathsf{UWB}\}$ simulated for the conventional optimization as well as using DSM with $\tilde{\nu}^\mathsf{UWB}=2.5^\circ$ for $T=\{1,2,4\}$.}
    \label{fig:matching}
\end{figure}

The impact of phase matching for DSM is illustrated in Fig. \ref{fig:matching} across various values of $T$. It is evident that for $T=1$ and $T=2$, phase matching yields negligible benefits. However, a significant improvement is observed when $T=4$. The previous 2dB performance loss compared to the conventional method is mitigated with the application of phase matching. This improvement is attributed to the ability to align interference constructively at the UWB-estimated point, when the phase values become precise enough. By selecting an appropriate combination of phase values, the destructive interference at this point can be minimized, leading to performance similar to the conventional method.
Remarkably, in some instances, the performance of DSM surpasses that of the conventional method. This phenomenon arises from the formulation of the model-based optimization, where the additional attenuation $A_m(\tau)$ is applied to all actively phase-shifting RIS elements after optimization. In other words, if the performance of both methods is the same after the optimization, the RIS configuration with more switched-on elements will perform worse. Consequently, the phase-aligned DSM solution achieves comparable quality regarding constructive interference but with fewer elements required for phase shifting, thereby enhancing channel quality. It is worth to note that phase matching can also yield benefits when utilizing the ASM. However, this is typically observed only in extreme scenarios, such as when the absolute Rx angle is very high or when $T$ is set to excessively high values, resulting in very fine control over the phase values.

\section{Experimental Results}
\begin{figure}
\centering
\includegraphics[width=.6\linewidth]{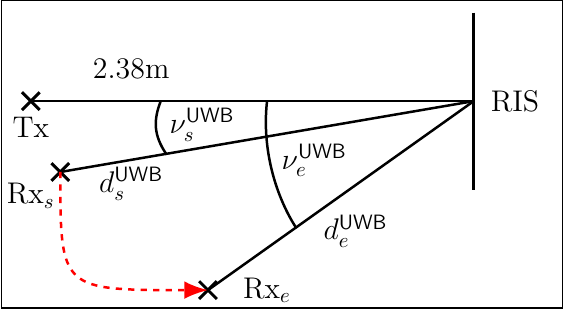}
\caption{Tracking scenario, where the Rx moves from Rx$_s$ to Rx$_e$ along the dashed arrow.}
\label{fig:traj}
\end{figure}

In the experimental study, the Tx antenna is situated directly in front of the RIS and positioned at a distance of 2.38m from it. Both, the Tx and Rx antenna are at the same elevation as the RIS center and oriented towards it. This alignment remains consistent even as the Rx antenna moves. The tracking scenario for the experimental evaluation is depicted in Fig. \ref{fig:traj}. In the beginning the Rx is positioned at starting point Rx$_s$ and then follows the trajectory indicated by the dashed arrow until it reaches the ending point Rx$_e$.

\subsection{Stationary Scenario}
In this section, we intend to analyze the precision of the UWB values. To this end, we first consider a stationary scenario. More precisely, we position the receiver at the trajectory's starting position Rx$_s$ (see Fig. \ref{fig:traj}) and optimize the RIS for all angles $\nu \in [-90,90]$ at a distance of 2.25m from the RIS center, effectively sweeping the environment, while simultaneously recording the UWB values.

\begin{figure}
        \centering
        \subfloat[UWB and corrected distances]{
            \includegraphics[width=.45\linewidth]{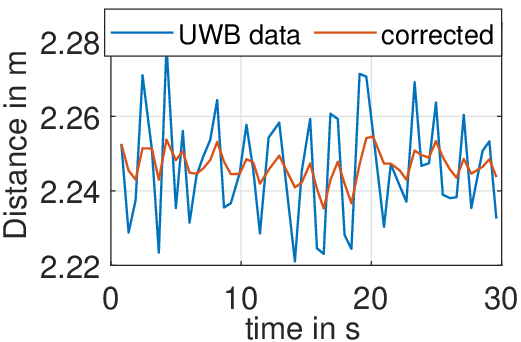}
            \label{subfig:A}
        }
        \subfloat[UWB and corrected angles]{
            \includegraphics[width=.45\linewidth]{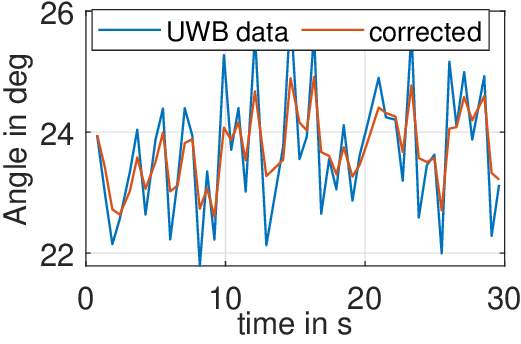}
            \label{subfig:B}
        }
        \caption{UWB and corrected values for the stationary case.}
        \label{fig:meas_stat}
\end{figure}

Fig. \ref{fig:meas_stat} displays recorded distances (a) and angle values (b) derived from UWB measurements alongside calculated corrected values for the stationary scenario.
The figures indicate that distances in (a) fluctuate within approximately a 6cm range, while angle values in (b) fluctuate within a range of around $4.2^\circ$.
Moreover, the figures demonstrate the effectiveness of the gradient-based method for determining corrected distance and angle values. Specifically, the corrected angle values now remain within a range of 2cm, resulting in a reduction of variations by approximately 66\%, while corrected angle values remain within $2.3^\circ$, decreasing the range by roughly 45\%.

\begin{figure}
\centering
\includegraphics[width=.9\linewidth]{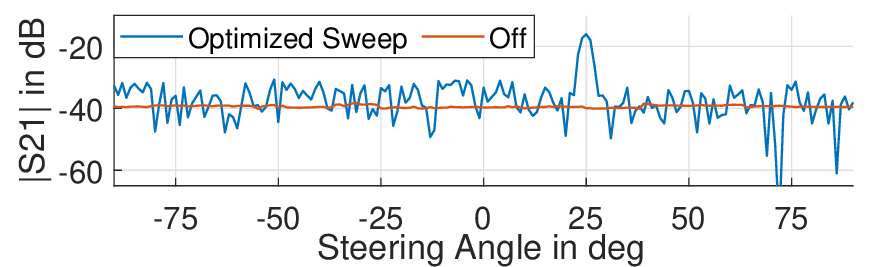}
\caption{RIS configuration sweeping all angles $\nu$ for the stationary scenario.}
\label{fig:sweep}
\end{figure}

Regarding the precision of the values, Fig. \ref{fig:sweep} depicts the amplitude of the S21 parameters resulting from the angle-sweep using the RIS configurations. In the figure, a distinct maximum can be seen at $25^\circ$. When compared to the mean of the UWB values depicted in Fig. \ref{fig:meas_stat} (b), an offset of around $1.3^\circ$ can be noticed.
Further, another important observation can be made in Fig. \ref{fig:sweep}. That is, deviations of the angle estimates by more then $\pm3^\circ$ of the real angle do not result in a performance benefit anymore. In fact, the peak is characterized by two significant local minima at $\pm5^\circ$, meaning that a deviation of more than $\pm3^\circ$ can also significantly reduce the expected performance. Given the fluctuations in Fig. \ref{fig:meas_stat} (b) in combination with the $1.3^\circ$ offset a misalignment of more than $\pm3$ becomes possible.

\subsection{Tracking Scenario}
Based on the observations from the stationary scenario, the ASM method emerges as the more promising technique for the tracking scenario. In addition, we aim to mitigate the observed fluctuations by expanding the angular beam splitting to generate 5 equally-split beams, originating from 5 horizontally-divided and equally-sized sub-surfaces. Furthermore, the angle-splitting factor is set to $\tilde{\nu}^\mathsf{UWB} = 7$. It's worth noting that increasing the number of beams for ASM doesn't add complexity to the Phase Matching procedure since all beams remain optimal if they share the same phase (refer to Fig. \ref{fig:phasematch}). Thus, given a UWB estimate $\nu^\mathsf{UWB}$ during the tracking scenario, 5 beams are generated towards $\nu^\mathsf{UWB}+[-3.5^\circ, -1.75^\circ , 0^\circ, 1.75^\circ, 3.5^\circ]$. Consequently, considering the worst-case scenario for the UWB angle estimation, which includes uncorrected fluctuations and the offset, the highest expected error $\nu^\mathsf{err} = (\frac{4.2}{2}+1.3^\circ) = 3.4 < 3.5$ still falls within the angular range of the split beams.

\begin{figure}
\centering
\includegraphics[width=.9\linewidth]{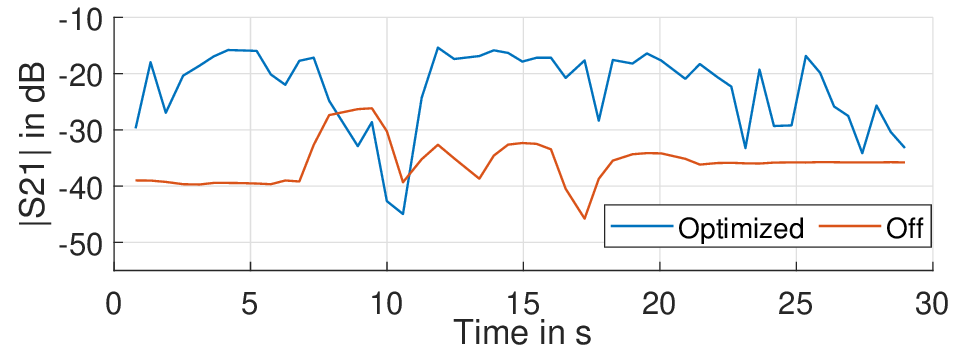}
\caption{Tracking scenario using the conventional method without beam splitting for uncorrected UWB values.}
\label{fig:m1}
\end{figure}

\begin{figure}
\centering
\includegraphics[width=.9\linewidth]{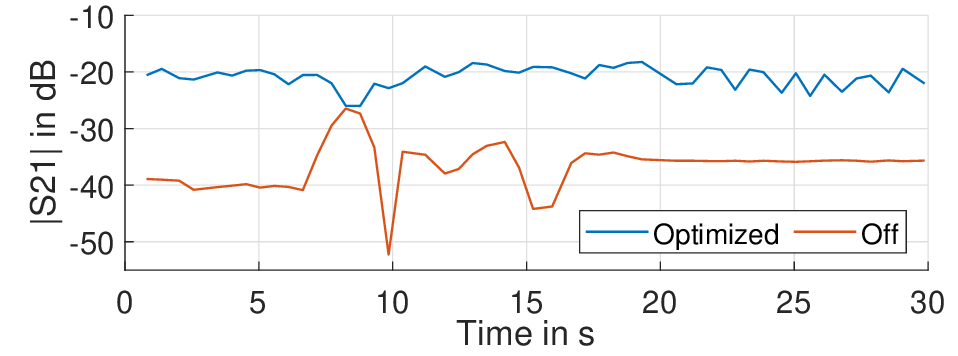}
\caption{Tracking scenario using ASM to generate 5 beams for uncorrected UWB values.}
\label{fig:m2}
\end{figure}

\begin{figure}
\centering
\includegraphics[width=.9\linewidth]{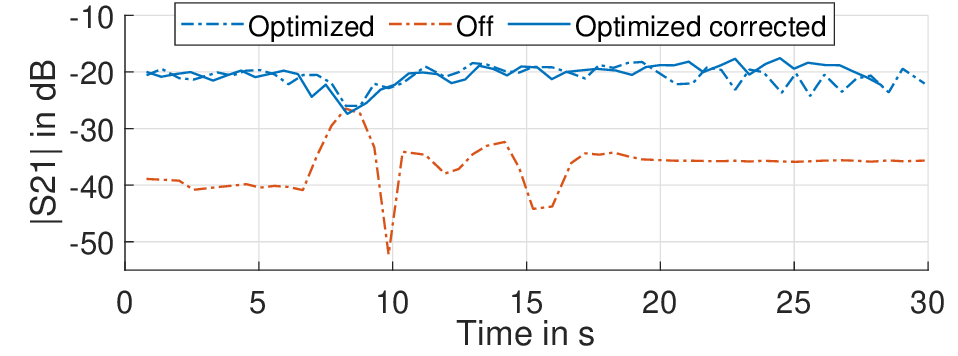}
\caption{Tracking scenario using ASM to generate 5 beams with and without corrected UWB values.}
\label{fig:m4}
\end{figure}

\begin{figure}
\centering
\includegraphics[width=.9\linewidth]{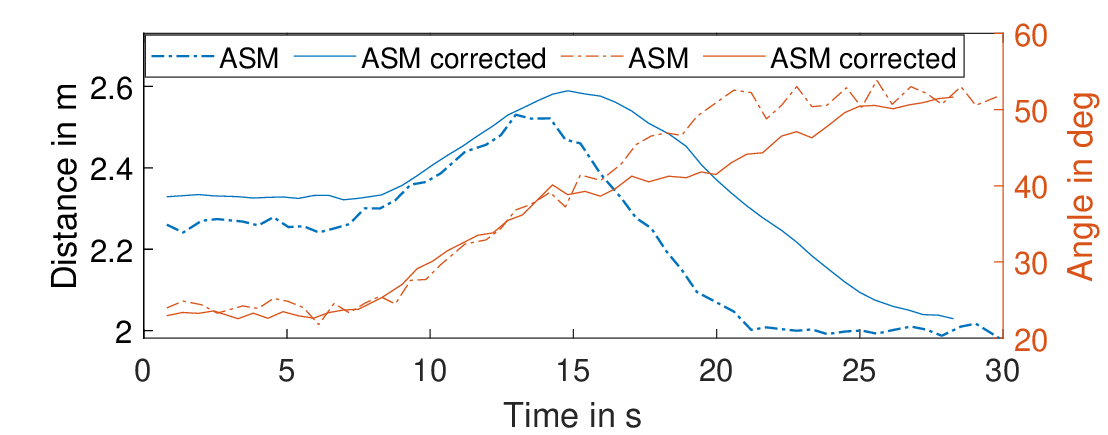}
\caption{Corrected and uncorrected UWB values.}
\label{fig:corrected}
\end{figure}

The effectiveness of the selected values is verified by comparing the tracking performance using uncorrected UWB values for the conventional method in Fig. \ref{fig:m1} with ASM in Fig. \ref{fig:m2}. In Fig. \ref{fig:m1}, it can be observed that the performance of the conventional method exhibits significant oscillations. This is due to the fact, that the RIS beam is following the uncorrected fluctuations of the UWB angle estimations (see Fig. \ref{fig:meas_stat} for the values of the stationary scenario). Additionally, it can be observed around the 8-second mark that the performance drops even below the off-state, where the RIS functions solely as a reflective plate. This observation aligns with the findings from the stationary scenario. Moreover, Fig. \ref{fig:m2} confirms that the countermeasures devised from the observations in the stationary scenario remain applicable in the tracking scenario. The figure illustrates that tracking performance is notably more consistent and reliable, as indicated by the smoother curve. Even around the 8-second mark, the performance remains better than the off-state configuration. Optimizing towards the corrected UWB values yields slightly improved consistency, particularly evident towards the end of the measurement, as depicted in Fig. \ref{fig:m4}. Even more performance benefits could be expected if the angle-splitting factor $\tilde{\nu}^\mathsf{UAB}$ would be reduced while utilizing the corrected values. This becomes feasible because the corrected values are fluctuating a lot less as seen in Fig. \ref{fig:corrected}. The figure demonstrates that both the corrected UWB distances and angles exhibit much smoother variations compared to their uncorrected counterparts, facilitating the creation of more densely split beams.

\section{Conclusion}

In this study, we equip a RIS prototype with an UWB localization system to enable real-time tracking of a moving receiver antenna. By leveraging UWB estimations, we can accurately determine the system's geometry, allowing us to employ model-based optimization techniques for configuring the RIS to maximize the received signal strength in real time. However, variations in the UWB estimates for the receiver's angle and distance values can lead to situations, where conventional optimization of the RIS configuration might result in the RIS reflected beam missing the receiver. To address this issue, two techniques that suppress the impact of the fluctuations in the UWB-estimated data are introduced. The first technique is the beam splitting method, which increases the probability of hitting the receiver with a beam by splitting the surface into sub-surfaces. These surfaces are then optimized to points around the current UBW-estimate. The second technique is the UWB estimation correction, which reduces the variation in the UWB estimates by leveraging the concept of the momentum-based gradient descent approach. Experimental measurements of stationary and mobile scenarios validate the performance-enhancing benefits of the proposed techniques. In fact, when both techniques are combined, the worst-case tracking performance is increased by a remarkable 17.5dB compared to the conventional method.
\bibliographystyle{IEEEtran}
\bibliography{bibliography}
\balance
\end{document}